\documentclass[aps,pra, twocolumn]{revtex4-1}
\usepackage{hyperref}
\hypersetup{colorlinks=true, urlcolor=blue}

\usepackage{newlfont}
\usepackage{amssymb}
\usepackage{amsfonts}
\usepackage{amsmath}
\usepackage{graphicx}
\usepackage{bm}

\usepackage{graphicx}
\usepackage{epsfig}
\usepackage{newlfont}
\usepackage{amssymb}
\usepackage{amsfonts}
\usepackage{amsmath}
\usepackage{graphicx}
\usepackage{bm}

\begin{document}

\title{Patterns of Genuine Multipartite Entanglement in Frustrated Quantum Spin Systems}

\author{Lavisha Jindal\(^{1,2}\), Ameya Deepak Rane\(^{3}\), Himadri Shekhar Dhar\(^4\), Aditi Sen(De)\(^5\), and Ujjwal Sen\(^5\)}

\affiliation{\(^1\)Department of Physics, St. Stephen's College, Delhi University, Delhi 110007, India\\ 
\(^2\)Department of Physics, Indian Institute of Technology Madras, Chennai 600036, India\\
\(^3\)Department of Physics, Indian Institute of Technology Delhi, New Delhi 110016, India\\
\(^4\)School of Physical Sciences, Jawaharlal Nehru University, New Delhi 110067, India\\
\(^5\)Harish-Chandra Research Institute, Chhatnag Road, Jhunsi, Allahabad 211019, India}

\begin{abstract}
  
We investigate the behavior of genuine multipartite entanglement of paradigmatic 
frustrated quantum spin systems.   
We consider six different spin models, whose frustration  ranges from being very high to very low.
We find that the highly frustrated quantum spin systems are near-maximally genuine multipartite entangled, while
those 
with low frustration do not possess a similar definite behavior with regard to their 
genuine multipartite entanglement.

\end{abstract}

\maketitle

\section{Introduction}
\label{intro}

\noindent Many-body systems are the ideal substrates to implement quantum information processing, ranging from state transmission to computational tasks \cite{maj_1,Fazio}. 
Successful implementation of such protocols in many-body systems crucially depends on the behavior of quantum correlations in such systems.
In the last few years, it has been realized that quantum correlations can be effectively used to predict cooperative phenomena in many-body systems \cite{maj_1,Fazio,Andreas-Nielsen}.
Further, quantum computation protocols, such as the cluster state quantum computation, have been proposed \cite{cluster1}, and experimentally realized in systems with a moderate number of qubits \cite{cluster2}. Developments in ultrcold atomic gases in optical lattices \cite{maj_2, opticallattice} and trapped ions \cite{iontrap} have enabled precise simulation of many-body Hamiltonians with remarkable control of system parameters. 

In recent years, frustrated spin systems have received a lot of attention due to its association with high-$T_c$ superconductivity \cite{hightc}, and the discovery of exotic frustrated phases \cite{exotic} (cf. \cite{moreexotic}).
Frustration arises from the failure to obtain an optimal configuration that simultaneously minimizes the energy in all interaction links, 
imposed by the Hamiltonian governing the system \cite{frus}. The inherent constraints in a frustrated system may arise due to lattice geometry, 
disorder, or simply from the nature of the interactions \cite{maj_1,maj_2}. There have been several attempts to characterize and study frustrated systems using quantum information 
concepts, which include entanglement area law (reviewed in \cite{area}) \cite{amader-frustu,frustu-ent}, fidelity \cite{frustu-fidel}, and quantum discord \cite{frustu-disc} (cf. \cite{ent-frustu}). 
Various aspects of entanglement have also been experimentally investigated in frustrated quantum spin systems \cite{frustu-expt}.

Most of the previous quantum information studies in frustrated systems are restricted to bipartite quantum correlation measures. One of the main obstacles in studying beyond 
bipartite quantum correlation measures is the incomputability and lack of definite characterization of quantum correlations in a multipartite scenario.
In the present paper, 
we investigate the behavior of genuine multipartite entanglement in frustrated systems by using a measure, defined in terms of a geometric quantity \cite{geom}, known as the \textit{generalized geometric measure} (GGM) \cite{GGM} (cf. \cite{GGM2}). 
The GGM ($\mathcal{G}$) can be efficiently computed for large system size and hence, is a very useful tool for measuring multipartite entanglement. 
We consider six prototype frustrated models, among which some are highly, some are moderately, and the rest are weakly frustrated. 
Highly frustrated systems considered are 
(i) the Ising gas  with antiferromagnetic (AFM) long range interactions and  (ii) the resonating valence bond (RVB) gas, which is a ground state of the AFM Heisenberg model with long range interaction. 
Systems with moderate frustration that are considered here are (iii) the 2D $J_1-J_2-J_3$ Heisenberg model with valence bond ground states and (iv) 
the Majumdar-Ghosh (MG) model. 
The 
(v) 2D nearest neighbor (NN) AFM Heisenberg model with additional diagonal interactions known as the Shastry-Sutherland (SS) model and the (vi) Ising ring with NN interactions (some of which are frustrated) are 
the weakly frustrated models considered.
%
In our prototype frustrated models, we observe that while the multiparty entanglement of highly frustrated quantum systems converge to 
the maximum value 
($\mathcal{G} \approx$ $\frac{1}{2}$)
for a large number of spins, systems with  low frustration does not exhibit a unique behavior with respect to multipartite entanglement. We find that moderate as well as low-frustrated systems can have high as well as low multipartite entanglement. 
It is known that entanglement area law fails to characterize highly frustrated systems.
On the contrary, we find here that genuine multipartite entanglement can characterize such systems. On the other hand, unlike genuine multipartite entanglement, weakly frustrated systems can have characteristic area laws, as shown in \cite{amader-frustu}. Our study indicates that multipartite entanglement along with the entanglement area law can potentially characterize the entire spectrum of frustrated spin systems.  

\begin{figure*}
\epsfig{figure= 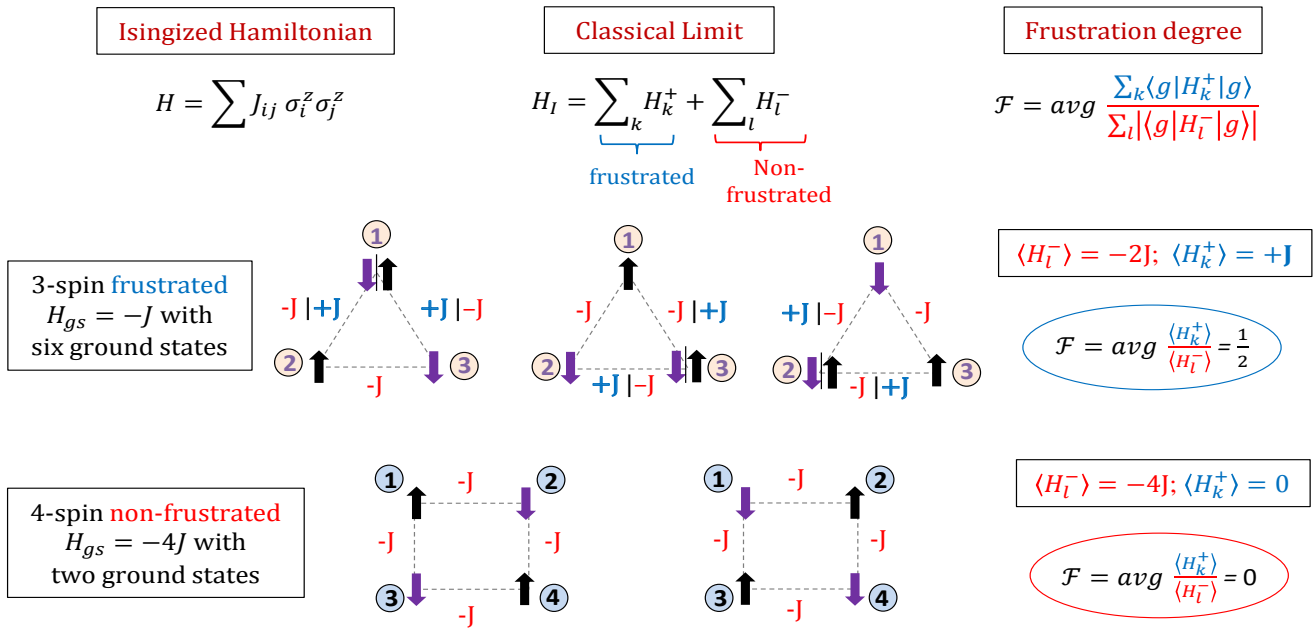, height=.35\textheight,width=.97\textwidth
}
\caption{(Color online.) Illustration of the definition of frustration degree in small spin systems. 
The first row contains a recollection of the general definition. The quantum Hamiltonian is 
first ``Isingized'', in the sense that we replace each of the terms of the Hamiltonian, 
irrespective of whether it is one-body, two-body, three-body, etc.,  by an Ising interaction term to obtain 
the ``Isingized Hamiltonian''. The values and signs of the interaction and field strengths are kept unchanged. This
Hamiltonian  is then transformed into the corresponding classical Hamiltonian by replacing the 
Pauli-\(\sigma_z\) matrices by the classical Ising variables (that take values \(\pm1\)). We then find the 
ground state configurations of this classical Hamiltonian. For every such ground state, we find out which term in the classical Hamiltonian has a positive (negative) contribution to the ground state energy, \(H_{gs}\), for that ground state. These ``positive'' (``negative'') terms of the Hamiltonian tend to increase (decrease) the \(H_{gs}\), and therefore
are incorporated into the ``frustrated'' (``non-frustrated'') parts of the Hamiltonian, for a given ground state. 
The ratio of the positive and the modulus of the negative contributions, averaged over all possible 
ground state configurations,  form the frustration degree of the quantum Hamiltonian.
In the second row, we consider, as an example, a frustrated antiferromagnetic 3-spin periodic linear quantum system, with six degenerate ground state configurations (each of the three schematic diagrams contain two possible configurations), both for the quantum as well as 
of the corresponding classical Hamiltonian. The presence of frustration is evident from the lack of a configuration of the classical (Ising) spins leading to a ground state energy of $H_{gs}=-3J$.
The contributions from each spin-pair interaction is shown next to the interaction line (light dashed). Each of these ground states are shown to have one positive-energy contributing (frustrating, shown in bold blue) term and two negative-energy contributing (non-frustrating, shown in red (non-bold font)) terms. The ratio of the positive and (modulus of) negative terms, averaged over the six possible ground states, is the frustration degree ($=\frac{1}{2}$), for the frustrated 3-spin quantum system. We next, in the third row, consider a 4-spin non-frustrated antiferromagnetic periodic linear quantum system with two degenerate ground states. The absence of frustration is evident in the figure from the presence of a classical spin configuration leading to the ground state energy $H_{gs}=-4J$. We observe that there is no positive-energy contributing (frustrating) term for both the ground state configurations. The corresponding frustration degree is thereby zero.}
\label{nf1}
\end{figure*}

The paper is arranged as follows. We quantify the genuine multipartite entanglement by using the GGM, which is 
defined in Sec. \ref{GGM}.
In Sec. \ref{frustu}, we present a quantification of the amount of frustration in a spin system, and also 
define a cooling mechanism that is required to investigate the genuine multisite entanglement of systems with 
degenerate ground states. In sections \ref{1}--\ref{6}, we present the behavior of  GGM  in the different 
frustrated systems. We conclude in Sec. \ref{concl}.

\section{Genuine Multipartite entanglement measure}
\label{GGM}

\noindent The generalized geometric measure quantifies the amount of genuine multipartite entanglement in a system \cite{GGM}.  
An \(N\)-party pure quantum state is said to be
genuinely \(N\)-party entangled, if it is not a product across
any bipartition. The simplest examples of genuine multipartite entangled states are the Greenberger-Horne-Zeilinger  \cite{GHZ} 
and W  \cite{W-state} states. The GGM 
of  an \(N\)-party pure quantum state \(|\psi\rangle\) is defined as
\begin{equation}
{\cal G} ( |\psi\rangle ) = 1 - \Lambda^2_{max} (|\psi\rangle ), 
\end{equation}
where  \(\Lambda_{max} (|\psi\rangle ) =\max | \langle \phi|\psi\rangle |\). The maximization is performed over all pure states \(|\phi\rangle\) that are not genuinely \(N\)-party entangled. 
By definition, the measure \({\cal G}\) vanishes for all pure multiparty states that are not genuine multiparty entangled and is non-zero for others. It has been shown that \({\cal G}\) does not increase under local operations and classical communication (LOCC). It turns out that the maximization involved in \({\cal G}\) can be simplified, and expressed in terms of Schmidt coefficients and hence it is possible to compute \({\cal G}\) for any state of an arbitrary number of parties in arbitrary dimensions.
The simplified form of the GGM of  \(|\psi\rangle\) reads
%
\begin{equation}
{\cal G}(|\psi\rangle) = 1 - \max \{\lambda^2_{{\cal A}: {\cal B}} | {\cal A} \cup {\cal B} = \{1,2,\ldots, N\}, {\cal A} \cap  {\cal B} = \emptyset\},
 \end{equation}
where $\lambda_{\mathcal{A}:\mathcal{B}}$ is the maximum Schmidt coefficient when we divide the \(N\)-party state \(|\psi\rangle\) into two subsystems, \({\cal A}\) and \({\cal B}\).

\section{Characterizing Frustration}
\label{frustu}
\noindent In this section, we will introduce a few tools that will be required to characterize the frustrated systems. In particular, we define the \textit{frustration degree}, to obtain an hierarchy among different frustrated systems. 

Typically ground states of frustrated systems are highly degenerate. To characterize the multipartite entanglement of such systems, we apply a cooling mechanism which 
is briefly described below.

\subsection{Frustration Degree}
\label{frustu-1}
\noindent The frustration of a spin system can be characterized using a measure, known  as the \emph{frustration degree} ($\mathcal{F}$) \cite{amader-frustu}.
Spin frustration is a classical concept (i.e., it is clearly defined for classical spin systems) and thus connected to a classical limit of the quantum Hamiltonian \cite{maj_1, maj_2}.
The characterization of frustration degree is based on 
performing a classical limit (Ising limit) of the possible quantum system.
For a quantum system with Hamiltonian $H$, 
a classical limit,
$H_I$ can be obtained by replacing the one-body, two-body, etc. terms, which are written in terms of Pauli operators ($\vec{\sigma}=(\sigma_x,\sigma_y,\sigma_z)$) with their 
Ising 
counterparts. If we take $|g\rangle$ to be a ground state of the system, in the classical limit, the Hamiltonian $H_I$ can be separated into  $H_i^{+}$ and $H_j^{-}$,
which, respectively, contain terms that contribute a positive or a non-positive value to the total energy corresponding to $|g\rangle$.
From the classical concept of frustration, it is evident that the positive terms contribute to frustration in the system by trying to increase the ground state energy, whereas the non-positive terms enable non-frustrated ground state configurations by attempting to lower the same
(see Fig. \ref{nf1} for an illustration).
Hence, $H_I=\sum_k H_k^{+} + \sum_l H_l^{-}$. The \emph{frustration degree} of the spin system, with Hamiltonian $H$, can then be defined as 
\begin{equation}
\mathcal{F}= \textrm{avg}~ \frac{\sum_k \langle g|H_k^{+}|g\rangle}{\sum_l |\langle g|H_l^{-}|g\rangle|},
\label{frudeg}
\end{equation}
where ``avg" denotes the average over all possible ground states, $|g\rangle$. 
Hence, the frustration degree is the ratio of the positive value contributing (frustrating) terms and the non-positive value contributing (non-frustrating) terms, averaged over all possible ground state configurations.
The value of $\mathcal{F}$ ranges from $0$ to $1$, 
and the higher the value of $\mathcal{F}$, stronger is the frustration in the system described by it.
For non-frustrated systems, the value of $\mathcal{F}$ is zero as there is no positive value (frustration) contributing term in the Hamiltonian for any ground state configuration (see Fig. \ref{nf1}). 

\subsection{Cooling Mechanism}
\label{frustu-2}

A typical aspect of frustrated systems is the 
highly degenerate ground state manifold.
To investigate the entanglement properties of the system, we wish to probe its low-temperature states and to that end, we apply the
prescription of a cooling mechanism
to obtain a fixed \emph{cooled} system state from the highly degenerate ground state space \cite{amader-frustu}.
The cooling method for a Hamiltonian of \(N\) quantum spin-\(1/2\) particles can be enumerated in the following steps: 
1) We consider the initial state to be a fully separable state of an \(N\)-qubit system, given by
$
|\Phi\rangle_{in} = |\psi\rangle_1 |\psi\rangle_2 \ldots |\psi\rangle_N.
$
2) We project the initial product state \(|\Phi\rangle_{in}\) onto the ground state manifold of the frustrated system under consideration. Hence, the final projected state reads
$
|\Phi\rangle_F = (\sum_i |\Gamma\rangle_i \langle \Gamma|) |\Phi\rangle_{in},
$
where \(|\Gamma\rangle_i\)'s are the degenerate ground state eigenstates of the frustrated system. 
This cooling mechanism is akin to evaporative cooling in cold gas systems and quenching in many-body physics.
The entanglement properties of the frustrated system under consideration can be investigated 
by using $|\Phi\rangle_F$.
However,
the final state depends on the choice of the initial state \(|\Phi\rangle_{in}\). 
Therefore to remove the dependence on the initial choice of states, we optimize over all possible initial states. Hence, to characterize the genuine multipartite entanglement of the 
system described the Hamiltonian $H_I$,
we define the GGM of the final state $|\Phi\rangle_{F}$ as
\begin{eqnarray}
{\cal G}(|\Phi\rangle_{F}) = \max_{|\Phi\rangle_{in}}\left(\sum_i |\Gamma\rangle \langle \Gamma|_i\right) |\Phi\rangle_{in}.
\end{eqnarray}


\section{Pattern of GGM for different frustrated models}
\label{frust}

\noindent We consider the six prototype frustrated models.
These prototypes are known to range from highly frustrated systems to ones with very low frustration. 
It is to be stressed here that the qualitative intensity of the frustration in these systems is independent 
of the parameter used to quantify frustration. 
The use of frustration degree is limited to establish a numerical value of the amount of 
frustration in these frustrated systems.
The frustration degrees 
for these models range from near 0 to 1. 
We address the question whether multipartite entanglement can be a useful tool to study frustrated models.

\subsection{The Ising Gas Model}
\label{1}

\noindent Let us first consider the highly frustrated Ising model with antiferromagnetic long range (LR) interactions -- the Ising gas model.  
The  Hamiltonian for the antiferromagnetic Ising gas, consisting of \(2 m\) spins, is described by 
\[H_{Ising} (\lambda) = (J/2m)(S - 2m \lambda)^2, \quad J>0 \]
where \(S = \sum_i \sigma_i^z\), $J>$ 0, and \(0\leq \lambda \leq 1\).
For arbitrary $J$, $\lambda$, and $m$, the frustration degree is
$
\mathcal{F}=\left(1+2\lambda-\lambda^2-1/m\right)/(1+\lambda^2).
$
For large $m$, and $\lambda=0$, $\mathcal{F}$ maximizes and goes to unity ($\mathcal{F} \approx$ 1).

Due to the symmetry of the system, without loss of generality, we can assume that the initial state for cooling is the same for all $2m$ spins and is given by
\(|\psi\rangle_{in} = \Pi_i (\alpha |0\rangle + \beta |1\rangle)_i\), where $|0\rangle$ and $|1\rangle$ are the eigenstates of the operator $\sigma_z$ with eigenvalues $\pm$1, respectively. After quenching, the final state is independent of the parameters \(\alpha\) and \(\beta\) of the initial state and hence the requirement of maximization of multipartite entanglement over possible initial states is not required. The final  state of $2m$ qubits after cooling, whose multipartite entanglement is to be found, is given by the (unnormalized) state
\begin{equation}
|\psi\rangle_{f} = \sum_{[[0,1]]} |0\rangle^{\otimes m(1 + \lambda)} \otimes  |1\rangle^{\otimes m(1 - \lambda)}.
\end{equation}
where $[[0,1]]$ stands for all possible combinations of $|0\rangle$ and $|1\rangle$ that respects the relative density $\frac{1+\lambda}{1-\lambda}$.
For maximum frustration, i.e. for, $\mathcal{F}$ = 1 (\(\lambda =0\)),   
the coooled state of the system has $2m\choose m$ terms in the above superposition. 

To obtain the multipartite entanglement using the GGM, 
we need to obtain the maximum of the highest eigenvalues of all possible bipartitions in the $2m$-spin system. The eigenvalue problem, 
for the case of the Ising gas model,
can be solved analytically. For a spin system with $2m$ spins, there are \(m\) distinct bipartitions of the form 
$n:2m-n$ ($ n \le m$). We now prove that the maximum eigenvalue across all bipartitions is given by the $2:2m-2$ partition.\\
 

\noindent\textbf{Theorem.} \emph{For the maximally frustrated Ising gas model with LR AFM interactions, the 
maximum  eigenvalue occurs for the $2:2m-2$ partition.} \\

\noindent\emph{Proof}: For an arbitrary bipartite partition of the total spins, $n:2m-n$ ($ n \le m$), the non-zero eigenvalues are ${n\choose j} {2m-n\choose m-j}/{2m\choose m}$, for $j$ = 0,1,...,$n$. It can be easily derived that the highest eigenvalue, for any $n:2m-n$ bipartition, corresponds to $j=n/2$, for $n$ even, and to $j=(n \pm 1)/2$, for $n$ odd. 

Let $\delta_n$ be the highest eigenvalue for the partiton $n:2m-n$,
where $ n \le m$.
For an ``even partition'', i.e. for  $n=i$ with $i$  even, and the consecutive ``odd partition'' with $n=i+1$ (such that $i+1 \le m$),
let $\chi_{e/o}$ denote the ratio of their highest eigenvalues $\delta_n$, so that  
\begin{eqnarray}
\chi_{e/o}&=&\dfrac{\delta_i}{\delta_{i+1}}=\dfrac{{i\choose \frac{i}{2}}{2m-i \choose m-\frac{i}{2}}/{2m\choose m}}{{i+1\choose \frac{i+2}{2}} {2m-i-1 \choose m+1-\frac{1}{2}}/{2m\choose m}},\nonumber 
\end{eqnarray}
which simplifies to $~\chi_{e/o}= \frac{i+2}{i+1}$.
Therefore, $\delta_i > \delta_{i+1}$, for all even values of $i$ ($i+1 \le m$).
Now we consider two consecutive even partitions $n=i$ ($i$, even) and $n=i+2$, where $i+2 \le m$. The ratio of their highest eigenvalues, $\chi_{e2/e1}$, is given by 
\[
\chi_{e2/e1} = \dfrac{\delta_{i+2}}{\delta_{i}} = \dfrac{(i+2)(i+1)}{(2m-i-1)(2m-i)},
\] 
which gives us $~\chi_{e2/e1} < 1$ for all values of $m$ and even values of $i$ (provided, $i+2 \le m$). Hence, $\delta_i > \delta_{i+2}$. Moreover, \(\delta_2>\delta_1=\frac{1}{2}\) for all \(m\). Therefore, 
the maximum eigenvalue is obtained for the smallest even value of $n$. Hence the proof.
\hfill\(\blacksquare\)\\

The calculation of the GGM 
is straightforward after obtaining the result that the highest eigenvalue can be obtained by considering the $2:2m-2$ partition of the Ising gas of $2m$ spins. The maximum eigenvalue obtained is ${2\choose 1}{2m-2 \choose m-1}/{2m\choose m}=\frac{m}{2m-1}$. Hence, the GGM is given by 
\begin{equation}
\cal{G}=\frac{m-1}{2m-1}.
\end{equation} 

%
\begin{figure}[h!]
\begin{center}
\epsfig{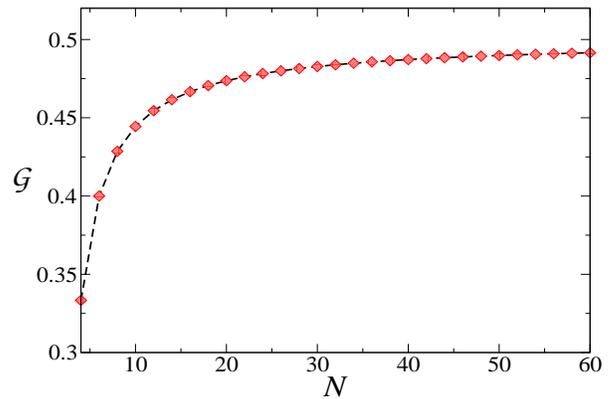}
\caption{(Color online.) GGM ($\mathcal{G}$) for the cooled state of the Ising gas. The horizontal axis represents the total number of spins, $N$ = \(2m\). We observe that $\mathcal{G}$ increases and finally, converges to $\frac{1}{2}$ as $N$ is increased. 
The vertical axis is dimensionless, while the horizontal one is measured in spins.}
\label{f1}
\end{center}
\end{figure}
%
The plot of $\cal{G}$ for different system size is given in Fig. \ref{f1}. We observe that $\mathcal{G}$ increases with increasing $m$. For large system size, 
$\mathcal{G} \rightarrow \frac{1}{2}$, and hence the ground state is maximally genuinely multipartite entangled for this maximally frustrated case.

\subsection{Resonating Valence Bond Gas}
\label{2}

\noindent Similar to the Ising gas, another highly frustrated system with the same frustration ratio ($\mathcal{F} \approx$ 1) is the Heisenberg antiferromagnet with long range interactions. The Hamiltonian is given by
\[
H_{RVB} = (J/2m) \sum_{i,j=1}^{m} \vec{\sigma}_i \cdot \vec{\sigma}_j, \quad J>0.
\]
The ground states of 
the Hamiltonian are all the states with total spin \(S=0\). 
To obtain the ground state, the total $2m$ spins are divided equally into two sets in which we mark $m$ spins as \emph{black} ($B$) and 
the remaining 
 as \emph{white} ($W$).
The ground state manifold of the Hamiltonian is obtained 
if one covers all $2m$ spins by singlets ($|\psi^-\rangle$) from  black to white spins \cite{overcomplete}.
%
%
Let us consider the initial state as 
\(|\psi\rangle_{in} = \Pi_{i=1}^{m} |\psi^{W}\rangle_i  \Pi_{i=1}^{m} |\psi^{B}\rangle_i\). 
The projection of the initial state in the \(S=0\) subspace leads to the final state, which is the equal
superposition of all the possible combinations of singlet coverings from black to white spins. 
The choice of the initial state is considered to be reasonable as every spin has two orthogonal states and 
there are two sublattices to be considered. 
Hence, the (unnormalized) cooled state is a resonating valence bond gas and can be written as
\begin{equation}
|\psi\rangle_{f}=\sum_k (|\psi^{-} \rangle_{BW} \cdots |\psi^{-} \rangle_{BW})_k ,
\end{equation}
where \(|\psi^{-} \rangle_{BW} \)
denotes the singlet shared from a black to a white spin. Each set of coverings ($k$) is a product of $m$ singlet coverings between pairs of black and white spins. The summation is over all such singlet configurations.
The total number of such possible coverings is $m!$.  

\begin{figure}[h!]
\begin{center}
\epsfig{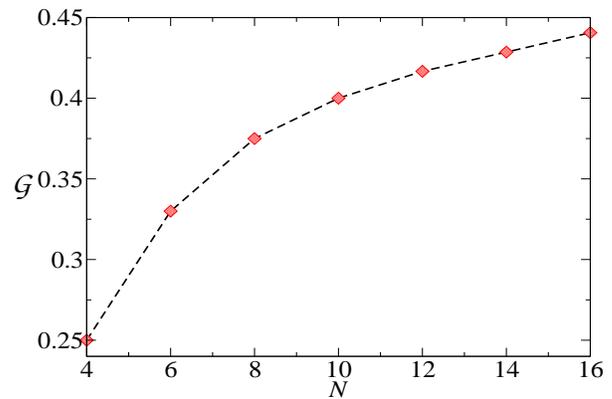}
\caption{(Color online.) GGM ($\mathcal{G}$) for the cooled state of the RVB gas, 
with respect to the total number of spins, \(N = 2m\). 
We observe that the $\mathcal{G}$ is an increasing function of $N$. Dimensions are the same as mentioned in Fig. \ref{f1}.}
\label{fig-RVB}
\end{center}
\end{figure}

Numerical simulation shows that the GGM
of the ground state, $|\psi\rangle_f$, increases with $m$ (see Fig. \ref{fig-RVB}).
For example, for $N(=2m)$ =16, the GGM is 0.44. 
From the figure (Fig. \ref{fig-RVB}), it is plausible that the GGM 
will possibly converge to some value close to the maximum GGM value ($\mathcal{G} \approx \frac{1}{2}$ ).

The discussions in the preceeding and current subsections indicate that 
highly frustrated systems, having maximum or near-maximum frustration degree, are maximally genuine multipartite entangled.

\subsection{2D \(J_1-J_2-J_3\) Heisenberg Model}
\label{3}

\noindent Let us consider a 2D square lattice containing 4$m^2$ spins, having \(J_1-J_2-J_3\) antiferromagnetic Heisenberg interactions with periodic boundary conditions. For certain specific values of the interaction strength, the entire spin lattice can be visualized in terms of $m^2$ square tetramers of neighboring spins called \emph{plaquettes} \cite{Kumar}. 

The Hamiltonian of such a model can be written as
\[
H_{\mathcal{H}}= J_1\sum_{\left<i,j\right>}\vec{\sigma}_i \cdot \vec{\sigma}_j+ ~J_2\sum_{\left<\left<i,j\right>\right>}\vec{\sigma}_i \cdot \vec{\sigma}_j+ ~J_3\sum_{\left<\left<\left<i,j\right>\right>\right>}\vec{\sigma}_i \cdot \vec{\sigma}_j,
\]
where 
$J_1$ is the interaction strength for the nearest neighbor and near diagonal spins on the same plaquette, $J_2$ is for the nearest neighbor and near diagonal spins on adjacent non-diagonal plaquettes, and $J_3$ is for the next-nearest neighbor and next-nearest diagonal (knight's move away) on adjacent non-diagonal plaquettes.
The summations in the Hamiltonian occur according to the permissible spin interactions.

The frustration degree of the 2D \(J_1-J_2-J_3\) Heisenberg model is relatively lower than the highly-frustrated systems considered in Sec. \ref{1} and Sec. \ref{2}. 
For systems with, e.g. $J_2$=$J_3$=0,
the frustration degree of the system is $\mathcal{F}$ = $\frac{1}{2}$. 

For specific choices of the relative spin interaction strengths in
the spin model, the ground state configuration is formed by a pair of either horizontal ($|hh\rangle$) or vertical singlets ($|vv\rangle$) in each of the $m^2$ plaquettes. The number of plaquettes containing $|hh\rangle$ or $|vv\rangle$ is determined by a pre-fixed density term, $\mathcal{D}$=$\dfrac{v}{m^2}$, where $v$ is the number of plaquettes that are $|vv\rangle$ for every covering. 
The cooling procedure can be implemented in the following way. The initial, unnormalized, state can be written as a product of the individual plaquette superpositions, $|\psi\rangle_{in}=\prod_k^{m^2} (\alpha|hh\rangle+\beta|vv\rangle)_k $, similar to the case for the Ising gas. The final state is independent of $\alpha$ and $\beta$. The quenched (unnormalized) final state is of the form
\begin{equation}
|\psi\rangle_{f} = \sum_{[[|hh\rangle,|vv\rangle]]} |vv\rangle^{\otimes \mathcal{D}m^2} \otimes  |hh\rangle^{\otimes (1-\mathcal{D})m^2}.
\end{equation}
The summation $[[|hh\rangle,|vv\rangle]]$ is over all possible combinations of $|vv\rangle$ and $|hh\rangle$, for fixed $\cal{D}$, which respects the relative density $\frac{\cal{D}}{1-\cal{D}}$. We consider $\mathcal{D}$ = 1/2, which is the state with the highest degeneracy, viz. $m^2\choose m^2/2$.  

We observe, from calculations, that the GGM 
decreases as the size of the system is increased. 
The system has very low genuine multipartite entanglement. For example, $\mathcal{G}$ is equal to 0.25, for $N$(=$4m^2$) = 4 (single plaquette) and reduces to 0.045, for $N$=16 (4 plaquettes). For non-square plaquette arrangement we again observe that $\mathcal{G}$ decreases with increase in the total number of spins. For example, we find that $\mathcal{G}$ is 0.1 and 0.0265 respectively for a lattice with $N$ equal to 8 and 24 spins, respectively.


\subsection{The Majumdar-Ghosh Model}
\label{4}

\noindent The Majumdar-Ghosh model \cite{MG} is a particular case of the AFM one-dimensional \(J_1-J_2\) model
with \(J_2/J_1 = \frac{1}{2}\).
It is interesting to note that in the entire \(J_1 -J_2\) --plane, the the ground states are exactly known only on the 
\(J_2/J_1 = \frac{1}{2}\) line. The Hamiltonian for the model is given by 
\begin{equation}
 H_{MG} =  J_1 \sum_{\langle ij\rangle} \vec{\sigma_i} \cdot \vec{\sigma_j} + \frac{J_1}{2}  \sum_{\langle\langle ij\rangle\rangle} \vec{\sigma_i} \cdot \vec{\sigma_j}, 
\end{equation}
where \(\langle ij\rangle\) and \(\langle\langle ij\rangle\rangle\) respectively denote the nearest neighbor and next-nearest neighbor interactions. Periodic boundary conditions are assumed.
The ground states 
of this model, for \(2m\) spins, for $\frac{J_2}{J_1}=\frac{1}{2}$ is doubly degenerate and is given by the following unnormalized state:
\[
 |G^{1,2}\rangle =  \Pi_{i=1}^{m} \frac{1}{\sqrt{2}}(|0_{2i} 1_{2i\pm 1} \rangle - |1_{2i} 0_{2i\pm 1}\rangle).
\]
The system is moderately frustrated with frustration degree $\cal{F}$=$\frac{1}{2}$.
The cooled state (unnormalized) in this case can be written as 
$
 |\psi\rangle_f = |G^1\rangle  + \alpha |G^2\rangle,
$
where \(\alpha\) can be any complex number.

Let us first consider the case when \(\alpha\) is real and maximize the GGM for all real \(\alpha\). The first step is to obtain the 
optimization involved in the calculation of $\cal{G}$ and second one is to maximize it over all possible initial states. 
Instead of the latter step, we maximize  over all \(\alpha\), and hence the obtained value of the GGM is, strictly speaking, an 
upper bound, in case the optimal \(\alpha\) cannot be reached by cooling a product state. 
It can be shown that the maximum eigenvalue 
is always obtained from 
the $2:2m-2$ bipartition.  Due to the symmetry of the state, 
there are two distinct $2:2m-2$ bipartitions, the corrsponding two distinct two-particle states 
being \(\rho_{12} = \mbox{tr}_{3\ldots 2m} (|\psi\rangle \langle \psi|_f) \) and 
\(\rho_{23} = \mbox{tr}_{1,4\ldots 2m}(|\psi\rangle \langle\psi|_f )\). 
The maximum eigenvalues of \(\rho_{12}\) and \(\rho_{23}\) can be analytically derived, and are respectively given by
\begin{eqnarray}
e_1 &=& \frac{2^{m+2} + 16 \alpha + 2^m \alpha^2}{4(2^{m} + 4\alpha + 2^m \alpha^2)}, \mbox{ and}\nonumber\\
e_2 &=& \frac{2^{m} + 16 \alpha + 2^{m+2} \alpha^2}{4(2^{m} + 4\alpha + 2^m \alpha^2)} \nonumber.
\end{eqnarray} 
The optimal value corresponds to $\alpha=1$, for all $m$, and this corresponds to $e_1 = e_2$.

\begin{figure}[h!]
\label{fig-MG}
\begin{center}
\epsfig{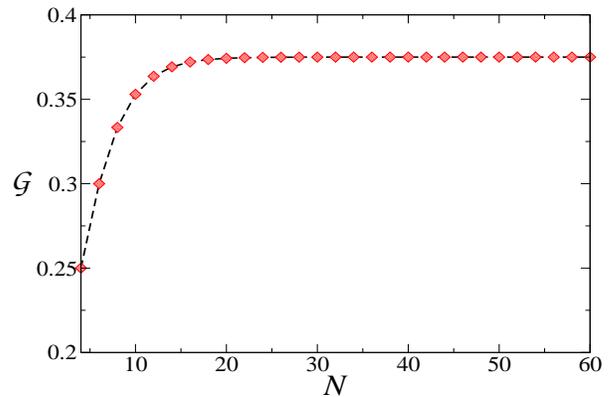}
\caption{(Color online) The GGM ($\mathcal{G}$) of the cooled state for the 
Majumdar-Ghosh model, with respect to the total number of spins, \(N = 2m\). We observe that $\mathcal{G}$ increases with $N$ and converges to $\mathcal{G} \approx \frac{3}{8}$ for large $N$. The dimensions of the axes are the same as mentioned in Fig. \ref{f1}. 
}
\label{fig-MG}
\end{center}
\end{figure} 

The GGM of the MG model can then be analytically calculated and is of the form
\begin{equation}
 {\cal G} = \frac{3}{(8 + 2^{4-m})},
\end{equation}
so that 
 for large systems, $\mathcal{G} \rightarrow \frac{3}{8}$ (as shown in Fig. \ref{fig-MG}).  

\begin{figure}[h!]
\label{fig-MG3D}
\begin{center}
\epsfig{figure= 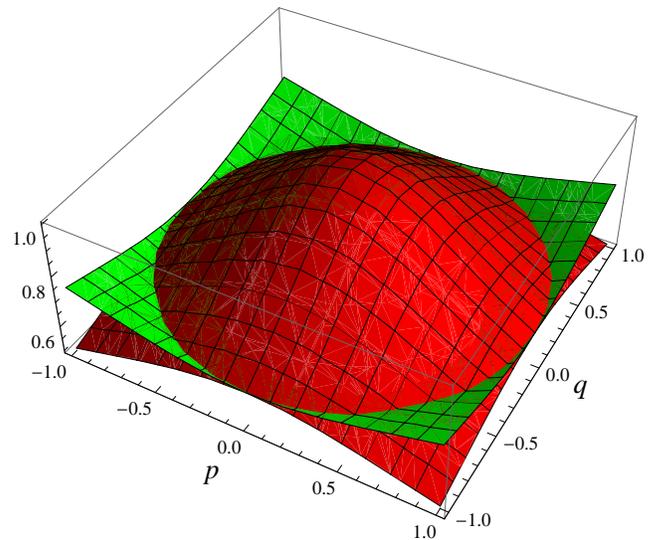, height=.3\textheight,width=0.5\textwidth}
\caption{(Color online) The maximum eigenvalues of \(\rho_{12}\) (red, dark) and \(\rho_{23}\) (green, light) of the MG model for \(m =40\). 
In grayscale, the maximum eigenvalue of \(\rho_{12}\) can be identified as the surface with a dome-shaped protrusion at the center of the base. The other surface corresponds to the maximum eigenvalue of \(\rho_{23}\).
The x- and y- axes represent \(p\) and \(q\), respectively. The z-axis represents the maximum eigenvalue of \(\rho_{12}\) and \(\rho_{23}\) for different values of $p$ and $q$. The eigenvalues are equal along the circle, $p^2+q^2$ = 1, which correspond to the points of maximum GGM. All axes are dimensionless.
}
\label{fig-MGD}
\end{center}
\end{figure} 

Let us now consider the more general situation when  \(\alpha\) is complex i.e., say, \(\alpha = p + i q\). We find that the maximum  \(\cal{G}\) is attained when the highest eigenvalues of \(\rho_{12}\) and \(\rho_{23}\) are equal, as in the case of real \(\alpha \). This occurs when
\(p^2 + q^2 =1\).
The optimization of eigenvalues over complex $\alpha$ is shown in Fig. \ref{fig-MGD}.  
For large system size, $\cal{G}$ again converges to \(\frac{3}{8}\) as in the case of real \(\alpha\).


\subsection{The Shastry-Sutherland model}
\label{5}

\noindent The Shastry-Sutherland model \cite{SS} is another example of two-dimensional AFM Heisenberg model on a square lattice, governed by the Hamiltonian
\[
H_{SS} =  J_1 \sum_{\langle ij\rangle} \vec{\sigma_i} \cdot \vec{\sigma_j} + ~J_2 \sum_{\left<\left<i,j\right>\right>} \vec{\sigma_i} \cdot \vec{\sigma_j},
\]
with periodic boundary conditions, where \(J_1\) is the nearest neighbor coupling ($\langle ij\rangle$) while \(J_2\) is the couplings for alternate diagonal ($\left<\left<i,j\right>\right>$). Due to the introduction of the coupling strength $J_2$, the system is frustrated. The model received a lot of attention after it was found that the material properties of the compound \(SrCu_2(BO_3)_2\) can be described by the SS spin Hamiltonian \cite{SS2}. 

The ground state 
of the model, for \(J_1/J_2 <\frac{1}{2}\), is a product of dimers along those diagonals whose coupling strength is \(J_2\). In the regime $J_1/J_2 <\frac{1}{2}$, the frustration degree of the system for large system size is $\cal{F}\approx \frac{1}{1+(1/2)(J_2/J_1)} < \frac{1}{2}$. 
Since the ground state of the SS model is product state over many bipartite cuts, the system has no multipartite entanglement, and hence the GGM is zero. Hence, we obtain a spin model with low frustration, $\cal{F} <$ 1/2, that has zero multipartite entanglement.

\subsection{Ising Ring with nearest neighbor interactions} 
\label{6}

\noindent Let us now consider a periodic 1D Ising spin model consisting of $2m$ spins with nearest neighbor (NN) interactions -- the Ising ring. 
For the Ising ring model, all but one nearest neighbor interactions 
are ferromagnetic ($-J$) while the remaining interaction is antiferromagnetic ($J$). The Hamiltonian of this model can be written as
\[
H_{Ring} = - J \sum_{i=1}^{2m-1} \sigma^z_i \sigma^z_{i+1} + J  \sigma_{2m}^z \sigma^z_{1},
\]
%
with \(J >0\). The frustration degree for the spin model is, $\cal{F}$ = $1/(2m-1)$. Similar to the SS model, this model is also very weakly frustrated. In the thermodynamic limit, the model does not have any frustration as $\mathcal{F} \rightarrow$ 0.

\begin{figure}[h!]
\label{fig-Isingnn}
\begin{center}
\epsfig{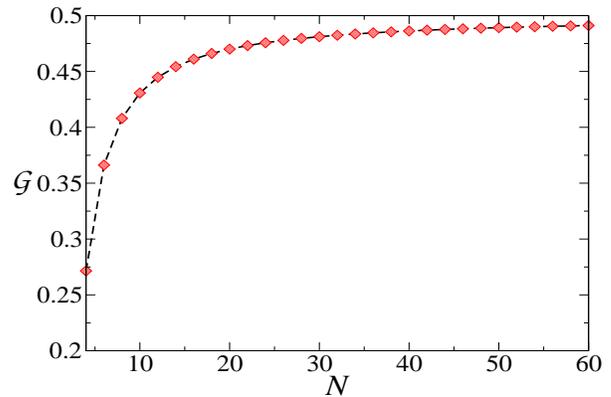}
\caption{(Color online.) GGM ($\mathcal{G}$) of the Ising Ring cooled state with respect to the total number of spins, \(N = 2m\). We observe that $\mathcal{G}$ increases with $N$ and is maximally multipartite entangled ($\mathcal{G}$ = 0.5) at the high $N$ limit. The dimensions of the axes are the same as mentioned in Fig. \ref{f1}. 
}
\label{fig-IR}
\end{center}
\end{figure}

The degeneracy in the ground state scales linearly with $m$, and is exactly \( 4 m\). If the initial unnormalized product state 
is chosen as \(|\psi\rangle_{in} = \Pi_i (|0\rangle + |1\rangle)_i\) (cf. \cite{ring}), the projected cooled state is given by 
\begin{eqnarray}
&&|\psi\rangle_f = \sum_{k =0}^{m-1} |0^{\otimes(2m-k)} 1^{\otimes k} \rangle + |1^{\otimes(2m-k)} 0^{\otimes k} \rangle  \nonumber \\
&&+|0^{\otimes(2m-k-1)} 1^{\otimes (k+1)} \rangle +  |0^{\otimes(k+1)} 1^{\otimes 2m -k -1} \rangle.
\end{eqnarray}
 
To calculate the multipartite entanglement using the GGM, one has to find the highest eigenvalue of each $n:2m-n$ bipartitions. It can be shown that the maximum eigenvalue that contributes to the GGM is from the \(2:2m-2\) bipartition. 
Therefore, the GGM is given by
\[
{\cal G} = \frac{3 m -1 - \sqrt{4 + m^2}}{4 m }. 
\] 
As evident from the Fig.\ref{fig-IR}, \({\cal G}\) reaches the maximum value \(\frac{1}{2}\),
for large $m$. 
The initial state that is chosen for this calculation is therefore an optimal one, atleast in the interesting region of large $m$.
The value of GGM obtained for the Ising ring is in contrast to the result obtained for the Shastry-Sutherland model in Section \ref{5}, where a weakly frustrated system led to zero GGM in the cooled state.

The discussions in the different subsections of this section lead us to the potential characterization of frustrated systems. Highly frustrated systems are highly genuine multiparty entangled. However, weakly frustrated systems can have a high or a low genuine multiparty entanglement. This is in direct contrast with the result obtained for area law in frustrated systems, where it is the weakly frustrated systems which follow the novel area law, while each of the strongly frustrated ones have their own area laws \cite{amader-frustu}.




\section{Conclusion}
\label{concl}
The study of quantum correlations in frustrated quantum spin systems is important for several aspects in many body physics and quantum information processing, including for example the realization of a 
quantum computer on a many-body platform. Manipulation of highly entangled many-body systems requires a clear understanding of the underlying quantum correlation characteristics that enable precise control and knowledge of the possible quantum applications in information theory and computation. Frustrated spin states have been simulated in experimental studies and offer exciting possibilities in harnessing the power of quantum correlations.

We have studied the behavior of multipartite entanglement for paradigmatic frustrated quantum spin systems, ranging from highly frustrated models to weakly frustrated ones. 
We use a  geometric measure  to compute the genuine multiparty entanglement present in such systems. 
The obtained pattern of multipartite entanglement is highly non-intuitive and does not evolve from known bipartite quantum correlation behavior. 
We observe that the highly frustrated systems are always near-maximally genuinely multiparty entangled. 
However systems with moderate or low frustration defy any specific behavior in terms of its genuine multipartite entanglement. 

%

\begin{acknowledgments}
LJ, ADR and HSD thank Harish-Chandra Research Institute (HRI) for hospitality and support during visits. HSD thanks University Grants Commission, India for financial support under the Senior Research Fellow scheme. The authors acknowledge the use of the cluster computation facility at HRI and the UGC-DSA computational facility at Jawaharlal Nehru University.
\end{acknowledgments}


\begin{thebibliography}{99}

\bibitem{maj_1} M. Lewenstein, A. Sanpera, V. Ahufinger, B. Damski, A. Sen(De), and U. Sen, Adv. in Phys. \textbf{56}, 243 (2007).

\bibitem{Fazio} L. Amico, R. Fazio, A. Osterloh, and V. Vedral,
Rev. Mod. Phys. \textbf{80}, 517 (2008).



\bibitem{Andreas-Nielsen} A. Osterloh, L. Amico, G. Falci, and R. Fazio, Nature \textbf{416}, 608 (2002); 
T.J. Osborne and M.A. Nielsen, Phys. Rev. A \textbf{66}, 032110 (2002).

\bibitem{cluster1} R. Raussendorf and H.-J. Briegel, Phys. Rev. Lett. \textbf{86}, 5188 (2001); R. Raussendorf, D.E. Browne, and H.J. Briegel, Phys. Rev. A, \textbf{68}, 022312 (2003); A.Y. Kitaev, Ann. Phys. \textbf{303}, 2 (2003);
M. van den Nest, A. Miyake, W. D{\"u}r, and H.J. Briegel, Phys. Rev. Lett. \textbf{97}, 150504 (2006); T.D. Ladd, F. Jelezko, R. Laflamme, Y. Nakamura, C. Monroe, and J.L. O’Brien, Nature \textbf{464}, 45 (2010).

\bibitem{cluster2} P. Walther, K.J. Resch, T. Rudolph, E. Schenck, H. Weinfurter, V. Vedral, M.
Aspelmeyer, and A. Zeilinger, Nature \textbf{434}, 169 (2005);
X.-C. Yao, T.-X. Wang, H.-Z. Chen, W.-B. Gao, A.G. Fowler, R. Raussendorf, Z.-B. Chen, N.-L. Liu, C.-Y. Lu, Y.-J. Deng, Y.-A. Chen, and J.-W. Pan, Nature \textbf{482}, 489 (2012).

\bibitem{maj_2} M. Lewenstein, A. Sanpera, and V. Ahufinger, \textit{Ultracold Atoms in Optical Lattices:Simulating quantum many-body systems} (Oxford Univ. Press, Oxford, 2012).

\bibitem{opticallattice} M. Greiner, O. Mandel, T. Esslinger, T.W. Hansch, and I. Bloch,
Nature \textbf{415}, 39 (2002); 
J.J. Garc{\'i}a-Ripoll, M.A. Martin-Delgado, and J.I. Cirac, Phys. Rev. Lett. \textbf{93}, 250405 (2004);
D. Jaksch and P. Zoller, 
Ann. Phys. \textbf{315}, 52 (2005);
J. Billy, 
V. Josse, Z. Zuo, A. Bernard, B. Hambrecht, 
P. Lugan, D. Cl{\'e}ment, L. Sanchez-Palencia, P. Bouyer, and A. Aspect, 
 Nature \textbf{453}, 891 (2008);
G. Roati,
C. D'Errico, L. Fallani, M. Fattori, C. Fort, M. Zaccanti, G. Modugno, M. Modugno, and M. Inguscio, 
\emph{ibid.} 895 (2008);
S. Trotzky, L. Pollet, F. Gerbier, U. Schnorrberger, I. Bloch, N.V. Prokof'ev, B. Svistunov, and M. Troyer, 
Nat. Phys. \textbf{6}, 998 (2010)


\bibitem{iontrap} D. Leibfried,
B. DeMarco, V. Meyer, M. Rowe, A. Ben-Kish, J. Britton, W.M. Itano, B. Jelenkovi{\'c}, C. Langer, T. Rosenband, and D.J. Wineland,   
Phys. Rev. Lett.
\textbf{89}, 247901 (2002);
D. Porras and J.I. Cirac, Phys. Rev. Lett. \textbf{92}, 207901 (2004);
H. H{\"a}ffner, W. H{\"a}nsel, C.F. Roos, J. Benhelm, D. Chek-al-kar, M. Chwalla, T. K{\"o}rber, U.D. Rapol, M. Riebe, P.O. Schmidt, C. Becher, O. G{\"u}hne, W. D{\"u}r, and R. Blatt, Nature \textbf{438}, 643 (2005);
A. Friedenauer, H. Schmitz, J.T. Glueckert, D. Porras, and
T. Schaetz, 
Nature Phys. \textbf{4}, 757 (2008);
R. Gerritsma, G. Kirchmair, F. Zähringer, E. Solano, R. Blatt, and C. F. Roos
Nature \textbf{463}, 68 (2010);


\bibitem{hightc} P.W. Anderson, Science \textbf{235}, 1196 (1987); P.A. Lee, N. Nagaosa, and X.-G. Wen, Rev. Mod. Phys. \textbf{78}, 17 (2006)


\bibitem{exotic}  Z.Y. Meng, T.C. Lang, S. Wessel, F.F. Assaad, and A. Muramatsu, Nature \textbf{464}, 847 (2010); A. Mulder, R. Ganesh, L. Capriotti, and A. Paramekanti, Phys. Rev. B \textbf{81}, 214419 (2010); A.F. Albuquerque, D. Schwandt, B. Het{\'e}nyi, S. Capponi, M. Mambrini, and A.M. L{\"a}uchli, Phys. Rev. B \textbf{84}, 024406 (2011); J. Reuther, P. W{\"o}lfle, R. Darradi, W. Brenig, M. Arlego, and J. Richter, Phys. Rev. B \textbf{83}, 064416 (2011);
C. N. Varney, K. Sun, V. Galitski, and M. Rigol, Phys. Rev. Lett. \textbf{107}, 077201 (2011);
B. K. Clark, D. A. Abanin, and S. L. Sondhi, Phys. Rev. Lett. \textbf{107}, 087204 (2011); Y. J. Yan, Z. Y. Li, T. Zhang, X. G. Luo, G. J. Ye, Z. J. Xiang, P. Cheng, L. J. Zou, and X. H. Chen, Phys. Rev. B \textbf{85}, 085102 (2012).   

\bibitem{moreexotic} M. Rasolt and Z. Tesanovi{\'c}, Rev. Mod. Phys. \textbf{64}, 709 (1992);
M. Sigrist and T.M. Rice, \emph{ibid.} \textbf{67}, 503 (1995);
%
R. Melzi,
P. Carretta, A. Lascialfari, M. Mambrini, M. Troyer, P. Millet, and F. Mila,
Phys. Rev. Lett. \textbf{85}, 1318 (2000); G. Misguich and C. Lhuillier, in \emph{Frustrated Spin Systems}, ed. H. T. Diep (World Scientific, Singapore, 2005);
F. Alet, A.M. Walczak, and M.P.A. Fisher, Physica A \textbf{369}, 122 (2006).















\bibitem{frus} G. Toulouse, Commun. Phys. \textbf{2}, 115 (1977); J. Vannimenus and G. Toulouse, J. Phys. C \textbf{10}, L537 (1977).

\bibitem{area} J. Eisert, M. Cramer, and M.B. Plenio, Rev. Mod. Phys. \textbf{82}, 277 (2010). 

\bibitem{amader-frustu} A. Sen(De), U. Sen, J. Dziarmaga, A. Sanpera, and M. Lewenstein,
Phys. Rev. Lett. \textbf{101}, 187202 (2008) 

\bibitem{frustu-ent} M. Haque, V. R. Chandra, and J. N. Bandyopadhyay, Phys. Rev. A \textbf{79}, 042317 (2009).

\bibitem{frustu-fidel} V. M. L. D. P. Goli, S. Sahoo, S. Ramasesha, and D. Sen, J. Phys.: Cond. Matt. \textbf{25}, 125603 (2013). 

\bibitem{frustu-disc} C-H. Fan, H-N. Xiong, Y. Huang, and Z. Sun, Quant. Inf. Comput. \textbf{13} 0452 (2013).


\bibitem{ent-frustu} C.M. Dawson and M.A. Nielsen, Phys. Rev. A \textbf{69}, 052316 (2004); P. Facchi, G. Florio, U. Marzolino, G. Parisi, and S. Pascazio, New J. Phys. \textbf{12}, 025015 (2010); S.M. Giampaolo, G. Gualdi, A. Monras, and F. Illuminati, Phys. Rev. Lett. \textbf{107}, 260602 (2011); U. Marzolino, S. M. Giampaolo, and F. Illuminati, arXiv:1302.6251.

%

\bibitem{frustu-expt} K. Kim, M.-S. Chang, S. Korenblit, R. Islam, E.E. Edwards, J.K. Freericks, G.-D. Lin, L.-M. Duan, and C. Monroe, Nature (London) \textbf{465}, 590 (2010); 
X.-S. Ma, B. Dakic, W. Naylor, A. Zeilinger, P. Walther, Nat. Phys. \textbf{7}, 399 (2011); K. R. K. Rao, A. Sen(De), U. Sen, H. Katiyar, T. S. Mahesh, A. Kumar, arXiv:1301.1834.




\bibitem{geom} A. Shimony, Ann. N.Y. Acad. Sci. \textbf{755}, 675 (1995); 
H. Barnum and N. Linden, J. Phys. A \textbf{34}, 6787 (2001); M. B. Plenio and V. Vedral, J. Phys. A {\bf 34}, 6997 (2001);
D.A. Meyer and N.R. Wallach, J. Math. Phys. {\bf 43}, 4273 (2002);
T.-C. Wei  and P. M. Goldbart, Phys. Rev. A {\bf 68}, 042307 (2003);
A. Osterloh and J. Siewert, Phys. Rev. A {\bf 72}, 012337 (2005);
A. Osterloh and J. Siewert, Int. J. Quant. Inf. {\bf 4}, 531 (2006);
R. Or\'{u}s, Phys. Rev. Lett. {\bf 100}, 130502 (2008);
R. Or\'{u}s, S. Dusuel, and J. Vidal, {\em ibid.} {\bf 101}, 025701 (2008);
R. Or\'{u}s, Phys. Rev. A {\bf 78}, 062332 (2008);
M. Balsone, F. Dell'Anno, S. De Siena, and F. Illuminatti, Phys. Rev. A \textbf{77}, 062304 (2008);
D. {\v Z}. Djokovi\'{c} and A. Osterloh, J. Math. Phys. {\bf 50}, 033509 (2009);
Q.-Q. Shi, R. Or\'{u}s, J. O. Fj\ae{}restad, and H.-Q. Zhou, New J. Phys. {\bf 12}, 025008 (2010);
R. Or\'{u}s and T.-C. Wei, Phys. Rev. B {\bf 82}, 155120 (2010).

\bibitem{GGM}  A. Sen(De) and U. Sen, Phys. Rev. A \textbf{81}, 012308 (2010).

\bibitem{GGM2} R. Prabhu, S. Pradhan, A. Sen(De), and U. Sen, Phys. Rev. A \textbf{84}, 042334 (2011);
M. N. Bera, R. Prabhu, A. Sen(De), and U. Sen, arXiv:1209.1523; A. Biswas, R. Prabhu, A. Sen(De), and U. Sen, arXiv:1211.3241; H. S. Dhar, A. Sen(De), and U. Sen, Phys. Rev. Lett. {\bf 111}, 070501 (2013); H. S. Dhar, A. Sen(De), and U. Sen, New J. Phys. \textbf{15}, 013043 (2013).

 
 






\bibitem{GHZ}D.M. Greenberger, M.A. Horne, and A. Zeilinger, in \emph{Bell's
 Theorem, Quantum Theory, and Conceptions of the Universe}, ed. M. Kafatos
 (Kluwer, Dordrecht, 1989) pp 69--72

\bibitem{W-state}A. Zeilinger, M.A. Horne, and D.M. Greenberger, in \emph{Proceedings of Squeezed States and Quantum Uncertainty}, edited by D. Han, Y.S. Kim, and W.W. Zachary (NASA Conference Publication No. 3135, NASA, Washington, DC, 1992).



\bibitem{overcomplete} Such RVB coverings make an overcomplete basis in the \(S=0\) subspace. 






\bibitem{Kumar} B. Kumar, Phys. Rev. B \textbf{66}, 024406 (2002); M. Mambrini, A. L{\"a}uchli, D. Poilblanc, and F. Mila, Phys. Rev. B \textbf{74}, 144422 (2006).

\bibitem{MG} C.K. Majumdar and D.K. Ghosh, J. Math. Phys. \textbf{10},
1388 (1969); C.K. Majumdar and D.K. Ghosh,  \emph{ibid.} \textbf{10}, 1399 (1969).


\bibitem{SS} B.S. Shastry and B. Sutherland, Physica (Amsterdam) \textbf{108B+C}, 1069 (1981).

\bibitem{SS2} S. Miyahara and K. Ueda, Phys. Rev. Lett. \textbf{82}, 3701 (1999).  

\bibitem{ring} H.J. Briegel and R. Raussendorf, Phys. Rev. Lett. \textbf{86}, 910 (2001).

\end{thebibliography}
\end{document}